\documentclass[10pt]{article}
\usepackage{sao2}
\usepackage{psfig,epsf}

\issue{2008, 63, 389--396}

\def\apj{Astrophys. J}

\def\apjs{Astrophys. J. Supp.}

\begin{document}
\markboth{M.L.~Khabibullina, O.V.~Verkhodanov}
	  {CATALOG OF RADIO GALAXIES WITH $z>0.3$. I: CONSTRUCTION OF THE SAMPLE}
\title{Catalog of Radio Galaxies with $z>0.3$. I:
Construction of the Sample}
\author{
M.L.~Khabibullina\inst{a}
\and O.V.~Verkhodanov\inst{a}
}
\institute{
\saoname
}
\date{October 2, 2008}{October 24, 2008}
\maketitle

\begin{abstract}
The procedure of the construction of a sample of distant ($z>0.3$)
radio galaxies using NED, SDSS, and CATS databases for further
application in statistical tests is described. The sample is
assumed to be cleaned from objects with quasar properties. Primary
statistical analysis of the list is performed and the regression
dependence of the spectral index on redshift is found.
\mbox{\hspace{1cm}}
\end{abstract}

\section{INTRODUCTION}
Radio galaxies, which are among the most powerful cosmic objects, allow us to study the evolution
 of matter and the dynamics of the expansion of the Universe at different cosmological epochs.

Investigations of these radio sources are linked to several
cosmological tests, which permit estimating the parameters and
evolutionary characteristics of the Universe. Radio galaxies are
identified with giant elliptical galaxies with absolute magnitudes
$M\sim-26$ and have black holes with masses on the of
$10^9M_{\sun}$ as their central engines. These facts allow the
radio galaxies to be used as tools for the study of the parameters
of the distribution of visible and dark matter, dynamics of the
Universe, and the history of the formation of its structure. Let
us point out various published methods developed to find
cosmological parameters from the observed characteristics of radio
galaxies
~\cite{vo_par1:Verkhodanov2_n,vo_par2:Verkhodanov2_n,vo_par3:Verkhodanov2_n},
such as:
\begin{itemize}
\item
 ``magnitude--redshift'' Hubble diagrams \linebreak
 \mbox{(``K--$z$'')}
   ~\cite{willott_k_z:Verkhodanov2_n};
\item
   the ''size--$z$'' diagram
   ~\cite{gurvits:Verkhodanov2_n,guerra:Verkhodanov2_n,jack_jan:Verkhodanov2_n};
\item
  the ``age--$z$'' diagram
   ~\cite{starob:Verkhodanov2_n};
\item
   the ``source number--flux' ($\log N-\log S$) distribution
   ~\cite{condon84:Verkhodanov2_n,condon89:Verkhodanov2_n};
\item
   search for clustering and determination of the clustering parameter
   ~\cite{blake_wall:Verkhodanov2_n};
\item
   search for gravitational lenses
   ~\cite{gravlens:Verkhodanov2_n};
\item
  investigation of the properties of the radio haloes of clusters of galaxies
    \cite{colafrancesko:Verkhodanov2_n};
\item
  cosmic background temperature variations caused by the interaction of the large-scale
structure (clusters of galaxies) at different redshifts with
cosmic microwave background and observed on small angular scales
(the Sunyaev--Zel'dovich effect due to inverse \linebreak Compton
scattering) \cite{sz:Verkhodanov2_n} and on large scales in
variable gravitational potential---the Sachs--Wolfe effect
  \cite{swe:Verkhodanov2_n}. Radio galaxies can be used as a tool for searching for clusters of galaxies~\cite{protoclust:Verkhodanov2_n};
  \item
 and a number of other methods (see, e.g., reviews
  \mbox{\cite{vo_par1:Verkhodanov2_n,vo_par2:Verkhodanov2_n,miley_debreuck:Verkhodanov2_n}}).
\end{itemize}

A striking feature of radio galaxies is that we observe them
virtually from their formation epoch, i.e., their radio sources
are so powerful that modern radio surveys have catalogued almost
all objects of this type~\cite{vo_par3:Verkhodanov2_n}. Such
objects can serve as good probes for the study of the formation of
clusters of galaxies. Thus \mbox{Venemans et
al.~\cite{protoclust:Verkhodanov2_n}} show that 75\% of the radio
galaxies with $z>2$ are associated with protoclusters. Based on
this fact, the above authors estimate that there are roughly about
3$\times$10$^{-8}$ of forming clusters per comoving Mpc$^3$ in the
redshift interval \mbox{$2<z<5.2$}  with an active radio source.
They also estimate the density of protoclusters within the same
redshift interval to be of the order of 6$\times$10$^{-6}$
Mpc$^{-3}$.

Identification of radio galaxies with giant elliptical galaxies
(gE) born as a result of mergers in the early epoch allows them to
be used not only to probe the formation of the large-scale
structure and test models of star formation. The (``K--$z$'')
diagram confirms the age of the old stellar population in gE-type
galaxies~\cite{willott_k_z:Verkhodanov2_n} and this fact can also
be used to compute the ages of these galaxies
\cite{rg_age:Verkhodanov2_n,rg_rc_age:Verkhodanov2_n,age_sys:Verkhodanov2_n}
and perform quick photometric redshift estimates for these objects
out to  $z\sim4$ \cite{verkh_phot_z:Verkhodanov2_n}.

Another important problem associated with the study of radio
galaxies is the origin of supermassive black holes. The estimates
of energy release in the central engines of radio galaxies at
redshifts \mbox{$4<z<5$}, when the Universe was slightly older
than one billion years (in terms of the $\Lambda$CDM model), yield
black-hole masses on the order of  $10^9 M_{\sun}$, whereas it
takes much more than one billion years for such black holes to
form \cite{debreuk_z:Verkhodanov2_n}. This paradox can be resolved
in terms of the hierarchical formation model, but the actual
number of such objects (black holes with masses on the order of
$10^9 M_{\sun}$ in the redshift interval \mbox{$4<z<5$)} in the
Universe is not known with certainty.

Distant radio galaxies are selected by a criterion based on the
spectral index of their continuum. To this end, lists of sources
with ultrasteep spectra are compiled, where the parameters needed
are the low--frequency data points on the continuum radio spectra.
Among such objects a high percentage of distant radio galaxies is
found~\mbox{\cite{dagkes:Verkhodanov2_n,blum_miley:Verkhodanov2_n,par_big3a:Verkhodanov2_n,par_big3b:Verkhodanov2_n,uss_list:Verkhodanov2_n}},
including the most distant objects with redshifts $z>4.5$:
$z=5.199$ \cite{debreuk_z:Verkhodanov2_n} and with $z=4.514$
\cite{kopylov_z:Verkhodanov2_n}.

Nowadays databases and identification techniques allow automating
search and identification of distant radio galaxies. A number of
authors~\mbox{\cite{deca_bul:Verkhodanov2_n,
deca_aat1:Verkhodanov2_n, deca_aat2:Verkhodanov2_n,
deca_azh:Verkhodanov2_n, deca_piter08:Verkhodanov2_n,
deca_bul2:Verkhodanov2_n, grg_piter08:Verkhodanov2_n}} applied
such approaches to  study radio sources using CATS~\footnote{\tt
http://cats.sao.ru}
\cite{cats:Verkhodanov2_n,cats2:Verkhodanov2_n} and
NED
\footnote{\tt http://nedwww.ipac.caltech.edu}
(NASA/IPAC Extragalactic Database) databases of astronomical catalogs and
SkyView virtual telescope~\footnote{\tt http://skyview.gsfc.nasa.gov}, which
allow performing morphological analysis of objects.
\begin{figure*}[tbp]
\centerline{\hbox{
\psfig{figure=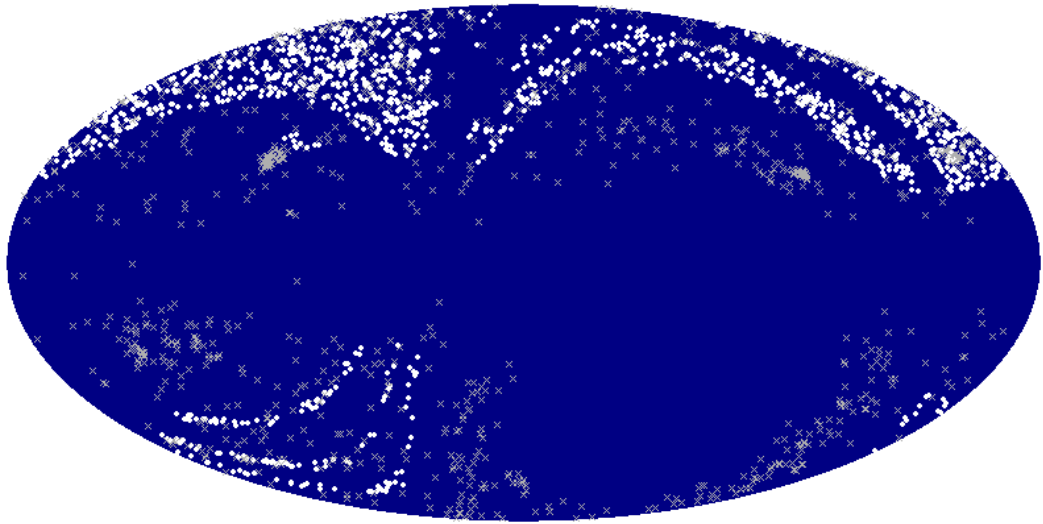,width=14cm}
}}
\caption{Sky positions of the selected radio sources in Galactic coordinates. The white circles and gray crosses indicate
the SDSS objects and all other sources, respectively.}
\label{f1:Verkhodanov2_n}
\end{figure*}

The catalog of distant radio galaxies can be used  not only to
perform just purely cosmological studies, but also carry out
statistical analyses of identification lists and of the
corresponding populations at different \mbox{wavelengths
\cite{irtex3:Verkhodanov2_n,irtex4:Verkhodanov2_n,balayan:Verkhodanov2_n,trushkin:Verkhodanov2_n};}
to search for and analyze the properties of subsamples of radio
galaxies~\cite{rg_gen:Verkhodanov2_n,rg_zen0:Verkhodanov2_n,rg_zen1:Verkhodanov2_n,rg_zen2:Verkhodanov2_n,grg_piter08:Verkhodanov2_n},
and simulate radio-astronomical surveys at \mbox{RATAN--600
\cite{compl:Verkhodanov2_n,rzf_cmb:Verkhodanov2_n,rzf_majorova:Verkhodanov2_n}.}
Organizing the databases and identifying the available
heterogeneous measurements with particular objects are also tasks
of great interest, and, as practical experience shows, such
compiled catalogs are popular among the astronomical community
\cite{cats2:Verkhodanov2_n,andernach:Verkhodanov2_n,vollmer:Verkhodanov2_n}.

Moreover, radio galaxies, as extended objects, may distort
microwave background and bias the estimates of the angular power
spectrum  \mbox{of
CMB~\cite{huffenberger:Verkhodanov2_n,grg_piter08:Verkhodanov2_n}.}

In this paper we report a catalog of radio galaxies with spectroscopic  \mbox{redshifts $z\ge0.3$} based on the data
from  the NED, SDSS\footnote{\tt http://www.sdss.org}\cite{sdss:Verkhodanov2_n}, CATS, and other databases and on an
analysis of the corresponding bibliography, which allowed us to discard  objects exhibiting quasar properties. There is
no yet a consensus  among the research community as to which radio galaxies should be considered  ``distant'', with
different authors using this term to denote objects with redshifts above  $0.3$, $0.5$, or $1.0$. Based on
observational data, one can set the boundary for distant radio galaxies at  $z\sim0.6-0.7$ (corresponding the age of
the Universe of about 6.5~Gyr), when, according to $\Lambda$CDM cosmology, the domination of dark energy (DE) began.
The lower redshift boundary can then be set at half this $z$ value. Such a choice permits the test to include
the population of galaxies after the beginning of the DE domination.

This is the first publication of a series of three papers
dedicated to the construction of the sample and statistical
analysis of the catalog of radio galaxies, which contains objects
with known spectroscopic~$z$ and available radio data and optical
photometry. In the future we plan to perform cosmological tests
based on the physical parameters of the objects of our catalog.

\section{THE CATALOG}

\subsection{Selection of Objects}

To build our primary list, we use the NED database, where we
select objects with the following parameters: redshift $z>0.3$ and
morphological type ``radio galaxy''. The initial list contained a
total of 3364 objects. This galaxy sample is polluted by objects
with incomplete data or objects with different properties. Our
next stage therefore consisted in cleaning the initial sample from
wrong sources. To this end, we select the following objects to be
removed from the initial list: (1) objects with photometrically
determined redshifts; (2) objects with quasar properties. We
performed an extensive scan of the literature including optical,
infrared, and radio surveys represented both in the NED and CATS
\cite{n2FGRS:Verkhodanov2_n,n2EX:Verkhodanov2_n,n2MASSi:Verkhodanov2_n,n2QZ:Verkhodanov2_n,n2SLAQ:Verkhodanov2_n,n3C:Verkhodanov2_n,n4C:Verkhodanov2_n,n6C:Verkhodanov2_n,n6dF:Verkhodanov2_n,n7C:Verkhodanov2_n,n87GB:Verkhodanov2_n,n8C:Verkhodanov2_n,ACS:Verkhodanov2_n,APMUKSbj:Verkhodanov2_n,B2:Verkhodanov2_n,B3:Verkhodanov2_n,CB:Verkhodanov2_n,CENSORS:Verkhodanov2_n,CFDF:Verkhodanov2_n,CM82:Verkhodanov2_n,CNOC2:Verkhodanov2_n,COINS:Verkhodanov2_n,CXOHDFN:Verkhodanov2_n,D0K:Verkhodanov2_n,DEEP2:Verkhodanov2_n,DMS:Verkhodanov2_n,dZP97:Verkhodanov2_n,EGS06:Verkhodanov2_n,F2M:Verkhodanov2_n,FBQS:Verkhodanov2_n,FCSS:Verkhodanov2_n,FIRST:Verkhodanov2_n,HB89:Verkhodanov2_n,HELLAS2XMM:Verkhodanov2_n,IRAS:Verkhodanov2_n,IRAS-L:Verkhodanov2_n,ISS:Verkhodanov2_n,KISSR:Verkhodanov2_n,LCRS:Verkhodanov2_n,MG1:Verkhodanov2_n,MOH2006:Verkhodanov2_n,MRC:Verkhodanov2_n,NEP:Verkhodanov2_n,NVSS:Verkhodanov2_n,OP:Verkhodanov2_n,PDFS:Verkhodanov2_n,PKS:Verkhodanov2_n,PMN:Verkhodanov2_n,RBS:Verkhodanov2_n,SDSS:Verkhodanov2_n,SGRS:Verkhodanov2_n,SSTXFLS:Verkhodanov2_n,SUMSS:Verkhodanov2_n,SWIRE:Verkhodanov2_n,SXDF:Verkhodanov2_n,TN:Verkhodanov2_n,TONS08:Verkhodanov2_n,TOOT:Verkhodanov2_n,TXS:Verkhodanov2_n,VIPS:Verkhodanov2_n,WARP:Verkhodanov2_n,WB92:Verkhodanov2_n}.
The complete
catalog is available from the CATS database\footnote{\tt ftp://cats.sao.ru/pub/CATS/RGLIST},
and CDS ftp\footnote{\tt ftp://cdsarc.u-strasbg.fr/pub/cats/J/other/AstBu/64.123},
CDS database\footnote{\tt http://cdsarc.u-strasbg.fr/cgi-bin/VizieR?-source=J/other/AstBu/64.123}.
Our sample also includes radio galaxies \linebreak investigated
within the framework of the \linebreak ``Big \mbox{Trio''
\cite{par_big3a:Verkhodanov2_n,par_big3b:Verkhodanov2_n,big_trio1:Verkhodanov2_n,big_trio2:Verkhodanov2_n,big_trio3:Verkhodanov2_n,big_trio4:Verkhodanov2_n,big_trio5:Verkhodanov2_n,big_trio6:Verkhodanov2_n}}
program, i.e.,  objects of the RC
catalog~\cite{rc1:Verkhodanov2_n,rc2:Verkhodanov2_n} with
published redshifts and measured photometric magnitudes.

As a result, we compiled a catalog containing 2442 objects for further studies.

Figure~1 shows the location of selected radio galaxies on the sky.

\subsection{Description of the Catalog}

The catalog of radio galaxies consists of three parts with each dedicated a separate paper.
Here we give the list of the objects with the names of the source catalogs (and, in the case of the 3C and
4C surveys, with the source names); equatorial coordinates (J2000.0); spectral indices for the 325, 1400, and
4850\,MHz frequencies, and spectroscopic redshifts. No spectral indices are given for galaxies with the fluxes
measured only at a single frequency (e.g.,  SDSS objects with the data from the  FIRST catalog \cite{FIRST:Verkhodanov2_n}). We compute the spectral indices based on the results of cross identification
in the CATS database with a  \mbox{200$''\times 200''$ identification box.} To remove spurious field radio
sources within the given box, we apply the technique of data analysis similar to that \mbox{described
by Verkhodanov et al.~\cite{deca_bul:Verkhodanov2_n,deca_bul2:Verkhodanov2_n}.}

The central idea of the
method consists in performing joint analysis of the data in the coordinate and spectral spaces in order to
identify likely identifications of particular radio sources at different radio frequencies. To this end,
we use the '{\it spg}' program~\cite{spg:Verkhodanov2_n} of the system of the reduction of continuum data
at RATAN-600 radio telescope. To represent the spectra  ($S(\nu)$) for the subsequent computation of
spectral indices we parameterize them by the following formula
$\lg\,S(\nu) = A + Bx + C f(x)$,
where $S$ is the flux in Jy;
$x$, the logarithm of frequency $\nu$ in MHz, and
$f(x)$ is one of the functions
$\exp(-x)$, \mbox{$\exp(x)$, or $x^2$.}

We collected the optical (DSS\,2) and radio images (mostly from the NVSS \cite{NVSS:Verkhodanov2_n}) of the
galaxies and their spectra into an atlas, which is available at
{\tt http://sed.sao.ru/$\sim$rita/Atlas\_RG.html}.

\subsection{Statistical Analysis of the Sample}

Figure~1 shows the locations of the selected galaxies in the sky.
The list as a whole is by no means complete or homogeneous, since
the sample contains objects brought from different catalogs and
located in mutually unconnected sky areas. The most complete and
homogeneous subsample of our list contains  SDSS objects, which
are indicated by white circles in Fig.\,1.  Among the SDSS radio
galaxies  (with the data from the NVSS and FIRST radio surveys)
there are many objects with small redshifts  \mbox{($z<0.5$)} and
small fluxes ($S<15$\,mJy), making the subsample in question stand
out qualitatively against other subsamples.

We nevertheless view the list reported here as a tentatively pure catalog of radio galaxies suitable for
testing cosmological relations. We call this list only tentatively pure, because we cannot say with
certainty that selected objects would not be found to exhibit  properties typical of another class.
The list remains open with both some objects to be possibly removed or new objects added.

The Table and Figs.~2--6 give the primary statistical parameters for the resulting catalog.
\begin{table}[tbp] \vspace{0.5cm}
\begin{center}
\caption{{\bf Table.} Parameters of the sample of the first part of the catalog of distant radio galaxies ($z>0.3$):
the median reshift, minimal, median, and maximum  1400~MHz flux in mJy, the minimum, median, and maximum spectral
index at 1400~MHz
}
\begin{tabular}{c|c|c|c|c|c|c}
$z_{med}$&$S_{min}$&$S_{med}$&$S_{max}$&$\alpha_{min}$&$\alpha_{med}$&$\alpha_{max}$ \\
\hline
 0.5    &  3.5     &  31.6   & 22720   &   -2.22       &    $-0.63$   &    $1.85 $    \\
\end{tabular}
\end{center}
\end{table}

Figure~2 shows the histogram of spectral indices at 1400\,MHz. We fitted the spectra of the sample mostly
by linear functions, resulting on  almost wavelength independent distribution. The median of the
distribution of the spectral indices for the entire sample is equal to --0.63. Figure~3 shows the
distribution of fluxes at the same frequencies~\footnote{We will report the fluxes in the third paper
of this series. The plots here are based on the data adopted from the  NVSS survey.}, and Fig.~4 shows the
distribution of redshifts.

\begin{figure}[!tbp]
\vspace{0.1cm}
\centerline{\hbox{
\psfig{figure=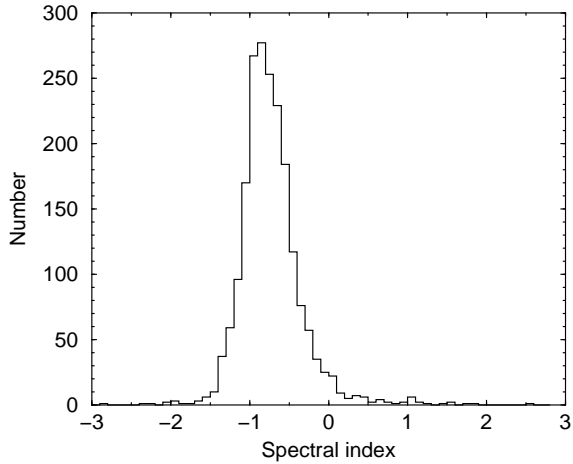,width=7.5cm,angle=-90}
}}
\caption{Histogram of spectral indices at  1400~MHz.}
\label{f2:Verkhodanov2_n}
\end{figure}

\begin{figure}[!tbp]
\centerline{\hbox{
\psfig{figure=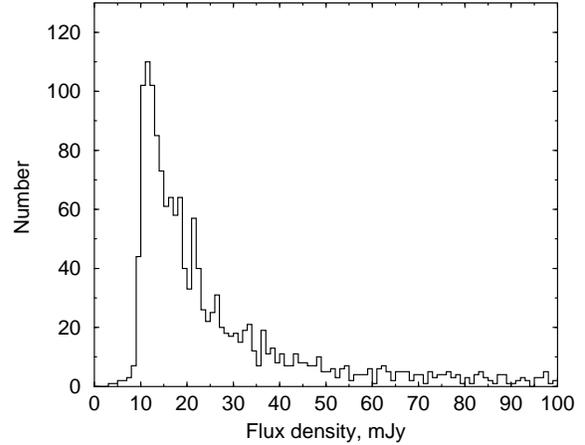,width=7.5cm,angle=-90}
}}
\caption{Histogram of the fluxes of radio galaxies according to
NVSS data.} \label{f3:Verkhodanov2_n}
\end{figure}

\begin{figure}[!tbp]
\centerline{\hbox{
\psfig{figure=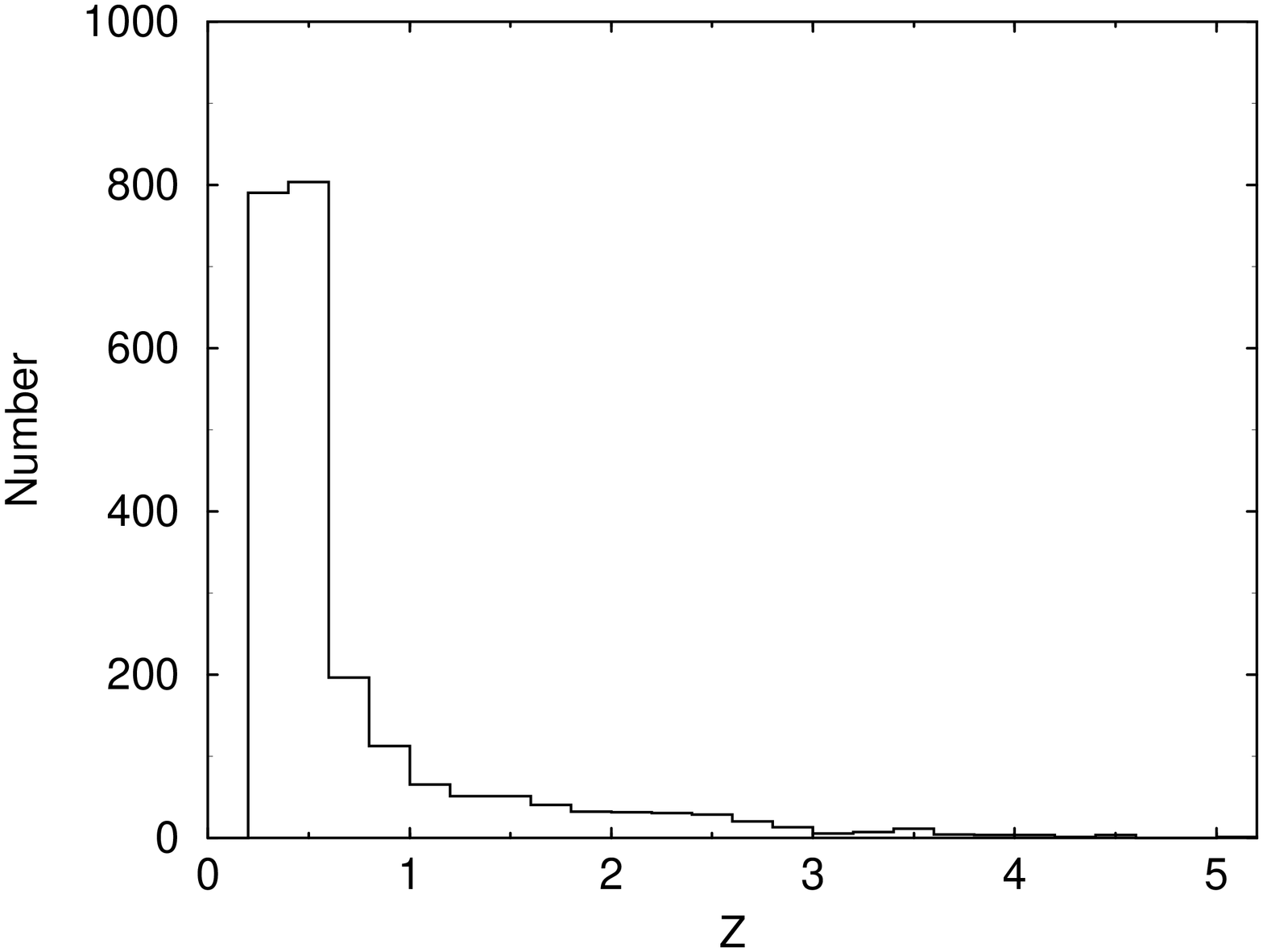,width=8cm,angle=-90}
}}
\caption{Histogram of  \mbox{redshifts.}}
\label{f4:Verkhodanov2_n}
\end{figure}

\begin{figure*}[!tbp]
\centerline{\hbox{
\psfig{figure=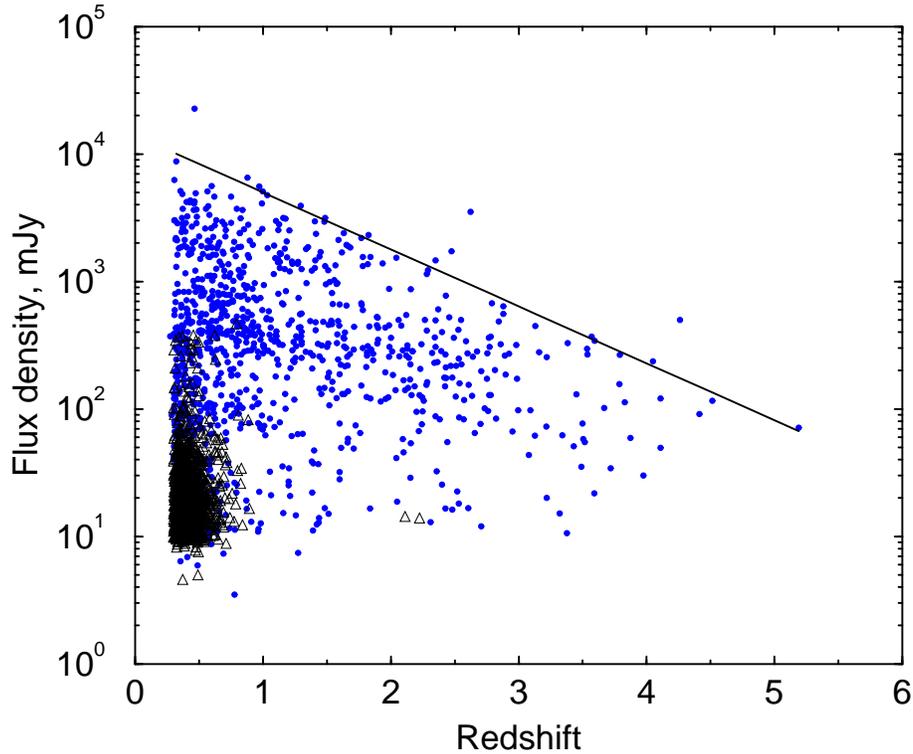,width=12cm,angle=-90}
}}
\caption{The ``observed flux--redshift'' diagram for 1400~MHz (the
NVSS catalog). The triangles and circles indicate the SDSS data
and objects of all other catalogs, respectively. The regression
fit is drawn for the maximum fluxes inside the bins with a
redshift step of $\Delta z = 0.5$. } \label{f5:Verkhodanov2_n}
\end{figure*}

Figures~5 and 6 show the ``flux--redshift'' and ``spectral
index--redshift'' diagrams, respectively.

The ``flux--redshift'' relation shows a well-defined upper flux
envelope  (at 1400~MHz, NVSS data), which gives the maximum
luminosities of observed radio galaxies at different $z$, and
apparently is due to the dynamics of cosmological expansion. We
fit linear regression  to the maximum values of the distribution
in $\Delta z=0.5$ wide bins. Our choice of the $\Delta z=0.5$ bin
width was determined by the following reasons: it must be (1)
small enough \mbox{($\Delta z\le1.0$)} to take the differential
properties of the population into account and (2) big enough
($\Delta z\ge0.1$) for the bins to contain sufficient number of
objects. We adopted $\Delta z=0.5$, which is, on the one hand,
halfway between the two extremes, and, on the other hand, ensures
that bins at $z>3$ contain a comparatively large number of
objects. The regression relation is given by the formula
\mbox{$\log S_{1400} =  p + rz$,} where $S_{1400}$ is the
1400\,MHz flux; $p=4.16$ is a constant intercept, and
\mbox{$r=-0.45$} is the slope of the line. The objects of this
sample are to be divided by morphological type and luminosity for
further analysis. Here we point out the especially powerful radio
galaxies with fluxes lie by more than half a magnitude above the
regression boundary: \mbox{3C 295}
\mbox{($z$=0.464)}~\cite{SDSS:Verkhodanov2_n}, \mbox{PKS 0742+10}
\mbox{($z$=2.624)}~\cite{labiano07:Verkhodanov2_n}, \mbox{8C
1435+635} \mbox{($z$=4.261)~\cite{spinrad95:Verkhodanov2_n}.} We
will analyze the properties of these objects in the third paper
dedicated to the catalog. We also do not plot the data point for
the radio galaxy VLA J123642+621331 ($z$=4.424)
\cite{richards98:Verkhodanov2_n}, which is included into the
catalog, but is absent in the NVSS list (because of its weak flux
(0.5~mJy)).

The ``spectral type--redshift'' diagram (Fig.\,6) shows a
well-defined trend---spectral index decreases with increasing $z$
and fits linear regression \linebreak relation $\alpha =  a + bz$,
where $a=-0.73\pm0.02$ and \linebreak \mbox{$b=-0.15\pm0.01$} are
the intercept and slope, respectively. Some objects exhibit peaks
near 1~GHz and at lower frequencies, which result in positive
spectral indices at low frequencies. In cases of negative
$\alpha$, with steep source spectra, we fitted the spectrum by a
linear relation. The spectral indices in the diagram is computed
at 1400\,MHz. Selection of objects by the steepness of their
spectra has been used extensively by a number of
authors~\mbox{\cite{blum_miley:Verkhodanov2_n,par_big3a:Verkhodanov2_n,par_big3b:Verkhodanov2_n,uss_list:Verkhodanov2_n},}
but we are the first to derive an analytical regression formula
for such a large sample of radio galaxies. We compute the
regression relation for the median of spectral indices in $z$ bins
with a bin width of $\Delta z =0.5$, which we adopt based on the
same considerations as in case of ``flux--redshift'' diagram (see
above), thereby making the inferred regression parameters stable
against the eventual future enlargement of the sample. The slope
can, in principle, be explained by two selection effects, although
physical causes cannot be ruled out:

(1) We adopt the lists of distant radio galaxies from the catalogs compiled by selecting objects with
steep spectra. The only way to avoid this effect is to use complete samples of radio galaxies with
measured redshifts, which are few.

\begin{figure*}[!tbp]
\centerline{\hbox{
\psfig{figure=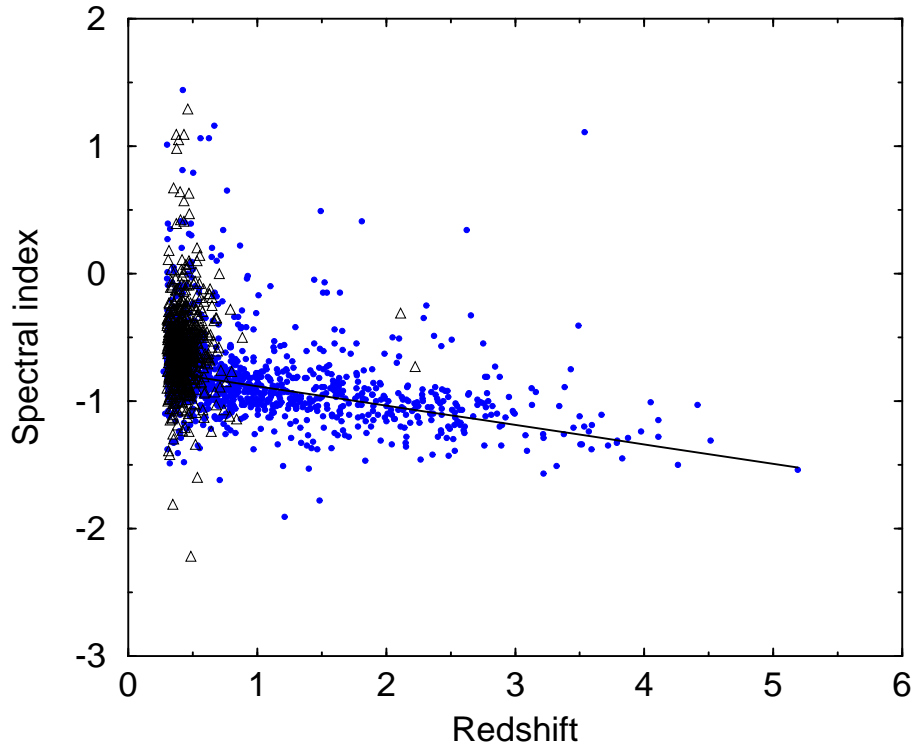,width=12cm,angle=-90}
}}
\caption{The ``spectral index--redshift'' diagram  (the spectral
indices are computed at 1400~MHz). The triangles and circles
indicate the SDSS data and objects of all other catalogs,
respectively. The regression fit is drawn for the median spectral
indices inside the bins with a redshift bin size of $\Delta z =
0.5$. }
\label{f6:Verkhodanov2_n}
\end{figure*}

(2) When we select the most powerful radio sources at large $z$ the radio flux is contributed by hot spots,
which have steep spectra. And the farther is the radio source, the more likely for it to be more powerful
compared to the surrounding sources and to have a steep radio spectrum. We plan to verify this fact in our
future studies by analyzing subsamples of radio galaxies of different luminosities. However, despite the scatter
of spectral indices, the analytical form of the regression can be used for preliminary selection of
radio galaxies and obtaining preliminary distance estimates, the regression yields a median estimate for
a large number of objects.

\section{CONCLUSIONS}

We report the first part of the catalog of radio galaxies with
$z>0.3$ with complete information on the coordinates of the
objects, their spectral indices at radio frequencies, and
redshifts. We constructed our catalog based on the lists of
objects from the  NED and CATS databases. Our sample includes
objects of the SDSS survey and of the ``Big Trio'' program. We
perform preliminary statistical analysis and compute the
distributions of spectral indices and fluxes and plot diagrams of
parameters as functions of  $z$. The catalog forms a basis for
further study of the objects and of the properties of distant
radio galaxies population. For our objects we construct the
spectra and the ``flux--redshift'' and ``spectral
index--redshift'' diagrams. On the former diagram we determine the
upper boundary of the dependence of flux on  $z$, which can be
described by the regression relation \mbox{$\log S = 4.16 -
0.45z$.} The latter diagram yields the regression relation
\mbox{$\alpha(z) = -0.73-0.15z$.}

\noindent
{\small
{\bf Acknowledgments}.
This research has made use of the NASA/IPAC Extragalactic Database
(NED) which is operated by the Jet Propulsion Laboratory,
California Institute of Technology, under contract with the
National Aeronautics and Space Administration. We also made use of
the CATS database of radio astronomical catalogs
\cite{cats:Verkhodanov2_n,cats2:Verkhodanov2_n} and
FADPS\footnote{\tt http://sed.sao.ru/$\sim$vo/fadps\_e.html}
radio-astronomical data reduction
system~\cite{fadps:Verkhodanov2_n,fadps2:Verkhodanov2_n}. This
work was supported by the Program of the Support of Leading
Scientific Schools in Russia (the School of S.\,M.\,Khaikin) and
the RFBR grant nos.\,09-02-00298 and 09-02-92659-Ind. O.V.V.
acknowledges partial support from the Russian Foundation for Basic
Research (project No\,07-02-01417-a) and the Foundation for the
Support for Russian Science (``Young Doctors of Science of the
Russian Academy of Science'').
}

\end{document}